\documentclass[prB,twocolumn,superscriptaddress,showpacs,amssymb]{revtex4}

\usepackage{psfrag,graphicx}
\usepackage{dcolumn}
\usepackage{bm}
\usepackage{amsfonts,amssymb,amsmath}        
\usepackage{braket}
\usepackage{color}
\usepackage{wasysym}
\usepackage{tikz}

\usepackage{hyperref}
\hypersetup{
bookmarks=false,
pdftitle={Microwave photonics}
anchorcolor=blue,
colorlinks=true,
citecolor=blue,
urlcolor=blue,
linkcolor=blue}
\usepackage{marginnote}
\usepackage[author=Juanjo]{pdfcomment}
\usepackage{xcolor}

\newcommand{\be}{\begin{equation}}
\newcommand{\ee}{\end{equation}}
\newcommand{\bq}{\begin{eqnarray}}
\newcommand{\eq}{\end{eqnarray}}

\newcommand{\rf}[1]{(\ref{#1})}

\newcommand{\bj}{\boldsymbol{j}}
\newcommand{\pp}{\boldsymbol{p}}
\newcommand{\da}{\dagger}
\newcommand{\bpm}{\begin{pmatrix}}
\newcommand{\epm}{\end{pmatrix}}
\newcommand{\noi}{\noindent}
\newcommand{\nn}{\nonumber}
\newcommand{\bs}{\boldsymbol}
\newcommand{\parti}{\tikz\draw[black,fill=black] (0,0) circle (.3ex);}
\newcommand{\Parti}{\tikz\draw[black,fill=black] (0,0) circle (.5ex);}
\newcommand{\hole}{\tikz\draw[black,fill=white] (0,0) circle (.3ex);}
\newcommand{\Hole}{\tikz\draw[black,fill=white] (0,0) circle (.5ex);}
\newcommand{\tgam}{\tilde{\gamma}}
\newcommand{\ta}{\tilde{a}}
\newcommand{\bxhat}{\hat{\bs{x}}}
\newcommand{\byhat}{\hat{\bs{y}}}
\newcommand{\bzhat}{\hat{\bs{z}}}

\begin{document}
\title{Nested Defects on the Boundary of Topological Superconductors} 

\author{James de Lisle}
 \affiliation{School of Physics and Astronomy, University of Leeds, Leeds, LS2 9JT, United Kingdom}
\author{Giandomenico Palumbo}
\affiliation{School of Physics and Astronomy, University of Leeds, Leeds, LS2 9JT, United Kingdom}
\affiliation{Institute for Theoretical Physics, Centre for Extreme Matter and Emergent Phenomena, Utrecht University, Leuvenlaan 4, 3584 CE Utrecht, The Netherlands}
\author{Adil A. Gangat}
\affiliation{School of Physics and Astronomy, University of Leeds, Leeds, LS2 9JT, United Kingdom}
\affiliation{ Department of Physics, National Taiwan University, Taipei 10617, Taiwan}
\author{Jiannis K. Pachos}
\affiliation{School of Physics and Astronomy, University of Leeds, Leeds, LS2 9JT, United Kingdom}

\date{\today}


\begin{abstract}
Helical Majorana edge states at the 2D boundaries of 3D topological superconductors can be gapped by a surface Zeeman field. Here we study the effect nested defects imprinted on the Zeeman field can have on the edge states. We demonstrate that depending on the configuration of the field we can induce dimensional reduction of gapless Majorana modes from 2D to 1D or quasi-0D at magnetic domain walls. We determine the nature of the Majorana localisation on these defects as a function of the magnitude and configuration of the Zeeman field. Finally, we observe a generalisation of the index theorem governing the number of gapless modes at the interface between topologically non-trivial systems with partial Chern numbers.

\end{abstract}

\maketitle
\section{Introduction}
In recent years interest in topological phases of matter has increased due to their fascinating properties and possible applications in spintronics and topological quantum computation. The latter is based on encoding quantum information in non-Abelian anyons and manipulating it by braiding the anyons with each other~\cite{pachos12}. A rich variety of non-Abelian anyons are predicted to emerge as quasiparticle excitations in fractional quantum Hall liquids. Unfortunately, the fragility of these systems complicates the creation, characterisation and manipulation of anyonic excitations. Hence, alternative media are sought that can support anyons.

Topological superconductors (TS) are an experimentally versatile medium comprising of free fermions with topological band structure. They are expected to support non-Abelian anyons in the form of Majorana zero energy modes localised at the core of vortices~\cite{Alicea10}. In fact, it is well known that TS support gapless Majorana modes localised at their boundaries that are decoupled from the bulk states of the system. Depending on the dimension of the TS they can support on their boundary 0D, 1D and 2D Majorana modes, like the Kitaev chain~\cite{Kitaev01}, the 2D p-wave TS in the class D~\cite{Pachos13} and the 3D TS in the class DIII, respectively~\cite{Finch15}. 

The study of gapless modes localised at defects in topological condensed matter and other systems has been long and fruitful~\cite{Volovik03,Santos11,Mandal15,Mandal15b,Pedro15,Ashai12,Shiozaki12}, from the prediction of solitons in polyacetylene~\cite{Su79,Jackiw76} and vortex fermions~\cite{Jackiw81,Chamon08} to Majorana modes trapped at vortex cores in the $p+ip$ superconductor~\cite{Read00}. Specific experimental implementations that give rise to Majorana zero modes have been proposed in superconducting-insulating heterostructures~\cite{Fu08} as well as semiconductor-superconductor heterostructures~\cite{Sau10,Alicea10,Lutchyn10,Oreg10}. Finally, it has been shown that 1D defects in the bulk of the 2D Kitaev honeycomb lattice can trap gapless Majorana modes~\cite{Lee07}.

In this paper we study the behaviour of Majorana modes at the boundary of a 3D TS in class DIII, which has both particle-hole (PH) and time-reversal (TR) symmetries. We initially consider a 3D model periodic in all three spatial dimensions that is characterised by a 3D winding number $\nu_\text{3D}\in\mathbb{Z}$. We then create a boundary by breaking the periodicity in the $z$ direction. The two disconnected surfaces naturally support gapless helical Majorana modes exponentially localised at the boundary. In order to gap these modes we introduce a TR-breaking Zeeman field thus effectively creating a 2D TS in class D. The topological phases at each boundary are unlike purely 2D TS as each of the two surfaces of the boundary is described by partial Chern numbers. As each surface can be manipulated independently it is intriguing to investigate how Majorana modes configure themselves around a variety of defects. To probe this we first introduce line defect in the Zeeman field between two regions with opposite field orientation. We numerically demonstrate that localised 1D Majorana modes live along these defect lines. In order to create 0D Majorana modes, we consider crossing 1D defect lines in the configuration of the effective Zeeman field. We demonstrate that quasi-0D Majorana modes are localised in the crossing points. We demonstrate our analysis for TS with $\nu_\text{3D}=1$ and $\nu_{\text{3D}}=2$. Overall we observe a generalisation of the well known index theorem \cite{Beri10} that determines the number of states at the interface between two system of differing bulk topological invariant. Motivated by this we demonstrate that the number of gapless Majorana modes is the difference in the partial contributions to the Chern number of the boundary system, either side of a magnetic domain wall.

In Section \ref{sec:model} we introduce the 3D DIII TS model. Section \ref{sec:2D} outlines the nature of the open boundaries and definitions and notation associated with the effective Zeeman field. Section \ref{sec:1D} introduces a Zeeman field configuration with 1D magnetic domain walls. We present numerical and analytical evidence showing this induces dimensional reduction of the surface states from 2D to 1D. Section \ref{sec:0D} introduces the `chessboard' configuration of effective Zeeman fields. We present numerical evidence showing the reduction of the Majorana modes from 2D to 0D. Finally, in section \ref{sec:conc} we summarise the preceding sections and suggest possible implementations using magnetic impurities or superconducting-to-magnetic material heterostructures.

\section{The Model}\label{sec:model}
\subsection{The Hamiltonian and its invariants}

The TS system we study is defined on a 3D lattice of size $N_{x}\times N_{y}\times N_{z}$, such that the position of a lattice site is given by  $\bj=(j_{x},j_{y},j_{z})$, where $1\leq j_{x,y,z}\leq N_{x,y,z}$. At each lattice site are two Dirac fermions $a_{1,\bj}$ and $a_{2,\bj}$ that form a unit cell. We consider the real space tight binding Hamiltonian given by~\cite{Finch15} 
\begin{eqnarray}\label{eqn:realham}
H &=&\sum_{\bj} \Big\{\sum_{k=1,2}\big[ 
	\mu a_{k,\bj}^\dagger a_{k,\bj} + \frac{1}{2}+ t a_{k,{\bj}}^\dagger a^{}_{k,\bj-\hat{\bs{x}}}\nn \\
	&+& 2t a_{k,\bj}^\dagger a_{k,\bj-\hat{\bs{x}}+\hat{\bs{z}}} + t a_{k,\bj}^\dagger a_{k,\bj+\hat{\bs{y}}} +t a_{k, \bj}^\dagger a_{k, \bj-\hat{\bs{y}}+\hat{\bs{z}}}\nn\\
 &+&\Delta a_{k,{\bj}}a_{k,\bj-\hat{\bs{x}}}+\Delta a_{k,\bj} a_{k,\bj-\hat{\bs{x}}+\hat{\bs{z}}}\nn\\
 &- & \Delta a_{k,\bj} a_{k,\bj+\hat{\bs{y}}} +  \Delta a_{k, \bj} a_{k, \bj-\hat{\bs{y}}+\hat{\bs{z}}}\big]\nn\\
 & - & 2i \Delta \big(a_{1,\bj} a_{2,\bj+\hat{\bs{y}}}+a_{2,\bj}a_{1,\bj+\hat{\bf y}}\big) \nn\\
 & + & 2i \Delta \big(a_{1,\bj}a_{1,\bj+\hat{\bs{x}}+\hat{\bf y}} - a_{2,\bj}a_{2,\bj+\hat{\bs{x}}+\hat{\bs{y}}} \big)+\mathrm{H.c.}\Big\} ,
\end{eqnarray}
where $\mu,\Delta,t\in\mathbb{R}$ are the chemical potential, pairing and tunnelling coefficients respectively, and $\hat{\bs{x}}$, $\hat{\bs{y}}$, $\hat{\bs{z}}$ are the unit vectors along the principal axes. 

Assuming translational invariance and periodic boundary conditions we can introduce the Fourier transform, $a_{k,\bj}=\sum_{\pp}e^{i\pp\cdot\bj}a_{k,\pp}$, and write Hamiltonian \rf{eqn:realham} in Bogoliubov-de Gennes form, $H=\sum_{\pp}\bs{\psi}_{\pp}^{\da}h(\pp)\bs{\psi}_{\pp}$, where $\pp\in[0,2\pi)\times[0,2\pi)\times[0,2\pi)$ and $\bs{\psi}_{\pp} = (ia_{1,\pp}-a^\dagger_{1,-\pp},a_{2,\pp} - ia^\dagger_{2,-\pp},ia_{1,\pp} + a^\dagger_{1,-\pp},a_{2,\pp}+ia^\dagger_{2,-\pp})^\text{T}/\sqrt{2}$, the spin-triplet basis. The kernel is given by
\begin{equation}
    h(\pp) =\bpm \epsilon(\pp)\mathbb{I}_{2} & \Theta(\pp) \\ \Theta(\pp)^{\da} & -\epsilon(\pp)\mathbb{I}_{2} \epm,
\end{equation}
where $\epsilon(\pp):T^{3}\mapsto\mathbb{R}$ is the insulating part with $\mathbb{I}_{2}$ the two-dimensional identity matrix and $\Theta(\pp)=i(\bs{d}(\pp)\cdot\bs{\sigma})\sigma_{y}$ is the spin-triplet pairing function with $\bs{\sigma}=(\sigma_x,\sigma_y,\sigma_z)$ the vector of Pauli matrices and $\bs{d}:T^{3}\mapsto \mathbb{R}^{3}$. The model possesses both PH and TR symmetries. This implies the kernel Hamiltonian satisfies
\begin{eqnarray}
C_\text{PH}h^{*}(\pp)C_\text{PH}^{\da}&=&-h(-\pp),\nn\\
C_\text{TR}h^{*}(\pp)C_\text{TR}^{\da}&=&h(-\pp),
\end{eqnarray}
where $C_\text{PH}=\mathbb{I}_{2}\otimes\sigma^{x}$ and a $C_\text{TR}=i\sigma^{y}\otimes\mathbb{I}_{2}$. As $C_\text{PH}^{2}=1$ and $C_\text{TR}^{2}=-1$ the symmetry class of the Hamiltonian is DIII. The model supports multiple gapped phases separated by gappless regions, where each phase is characterised by a bulk winding number \cite{Sato09}
\begin{equation}\label{eqn:3DWIND2}
    \nu_{\text{3D}} = \frac{1}{2}\sum_{\bs{d}(\pp^{*})=0}\text{sgn}\Big\{\epsilon(\pp^{*})\Big\}\text{sgn}\Big\{\text{det}\big[\textbf{J}[\bs{d}(\pp^{*})]\big]\Big\}
\end{equation}
where $\pp^{*}$ are the points in the Brillouin zone for which $\bs{d}(\pp^{*})=0$ and $\textbf{J}$ is the Jacobian. The model defined by \rf{eqn:realham} supports bulk winding numbers $\nu_{\text{3D}}=0,\pm1$ depending on the values of the couplings $\mu$, $\Delta$ and $t$. 

Consider the case where the system is in the topological phase, characterised by $\nu_{\text{3D}}=\pm1$. When we introduce open boundary conditions in the $z$ direction the system supports gapless surface states. Each surface of the boundary support a single Dirac cone that crosses zero energy at some point in the Brillouin zone; each surface state has linear dispersion. More generally, it was found \cite{Finch15} that the number of Dirac cones at each surface is equal to $\nu_{\text{3D}}$ when $\nu_{\text{3D}}\neq0$.

One can introduce an energy splitting $\Delta E$ in the surface states (to be distinguished from the bulk superconducting gap) by applying an effective Zeeman field to the boundary surfaces. In this event, when one considers the states of both the top and bottom surfaces together, their effective Hamiltonian in the low energy limit is that of a 2D TS in the class D. Such a system has a Chern number $\nu_{\text{2D}}$ such that $\nu_{\text{2D}}=\nu_{\text{3D}}$~\cite{Finch15}. To evaluate $\nu_{\text{2D}}$ we consider the contributions from each boundary surface, the top (T) and the bottom (B). The states on each surface make contributions $\nu_{b}\in\frac{1}{2}\mathbb{Z}$, $b=T,B$, defined by
\begin{equation}\label{eqn:FRACWIND}
    \nu_{b}=-\sum_{K \text{ on } b}\frac{i}{2\pi}\int_{\overline{\text{BZ}}}d^{2}p \,\text{tr}\Big(P_{K,\bar{\pp}}\big[\partial_{p_{x}}P_{K,\bar{\pp}},\partial_{p_{y}}P_{K,\bar{\pp}}\big]\Big).
\end{equation}
where $K$ denotes a Dirac cone, $P_{K,\bar{\pp}}$ are the projectors on the states that define each cone and $\bar{\pp}=(p_{x},p_{y})\in\overline{\text{BZ}}$ is the two-dimensional Brillouin zone of the boundary~\cite{Finch15}. These \emph{partial Chern numbers} sum to the full Chern number of the boundary $\sum_{b}\nu_{b}=\nu_{\text{2D}}\in\mathbb{Z}$, that characterises the topological phase of the whole boundary. This notion of considering the boundary consisting of the top and bottom surfaces as one delocalised system is of key importance in this work. 

Given two TS with bulk invariants $\nu_{+}$ and $\nu_{-}$, it is known \cite{Beri10} that the number of gapless states $N$ at the interface between the two systems is given by 
\begin{equation}\label{eqn:index}
    N=|\nu_{+}-\nu_{-}|.
\end{equation}
A system with an open boundary is equivalent to having $\nu_{-}=0$ and as such the number of gapless edge states is simply $|\nu_{+}|$. The bulk models presented in this paper are consistent with this description.

\subsection{Eigenvalues, eigenvectors, and spectral fermions}

We can factorise \eqref{eqn:realham} into the form
\begin{equation}\label{eqn:REALSPIN}
    H=\Psi^{\da} \Lambda\Psi,\,\,\,\,\, \text{where}\,\,\,\,\, \Psi = \bigoplus_{\bj}\bpm a_{1,\bj}\\a^{\da}_{1,\bj}\\a_{2,\bj}\\a^{\da}_{2,\bj}\epm.
\end{equation}    
The kernel $\Lambda$ is a square matrix of dimension $\text{dim}(\Lambda)=4N_{x}N_{y}N_{z}$. The positive eigenvalues of $\Lambda$ are enumerated $E_{n}$, $0\leq n\leq \text{dim}(\Lambda)/2$, where $E_{1}\leq E_{2}\leq ... \leq E_{\text{dim}(\Lambda)/2}$. The eigenstates of $\Lambda$ are given by
\begin{equation}
    \ket{\kappa_{n}}=\sum_{\bj}\sum_{k=1,2}\sum_{\rho=\parti,\hole}\alpha_{\bj k\rho}\ket{\bj}\otimes\ket{k\rho},
\label{eqn:states}
\end{equation}
where $\ket{\bj}$ are the position basis states and $\ket{k\rho}$ are mode basis states given by 
\begin{align}
    \ket{1\Parti}&=(1,0,0,0),& \ket{1\Hole}&=(0,1,0,0),\nn\\
    \ket{2\Parti}&=(0,0,1,0),& \ket{2\Hole}&=(0,0,0,1).
\end{align}
In \eqref{eqn:states} the complex number $\alpha_{k\rho\bj}$ is the amplitude of the $a_{k}$ ($\rho=\Parti$) or $a_{k}^{\da}$ ($\rho=\Hole$) fermionic mode at the site $\bj$. Particle-hole symmetry dictates that for every eigenstate $\ket{\kappa_{n}}$ with eigenvalue $E_{n}$ there exists a conjugate state $\ket{\kappa_{-n}}$ with eigenvalue $E_{-n}=-E_{n}$ such that $\ket{\kappa_{-n}}=C_{\text{PH}}\ket{\kappa_{n}}$ where $C_{\text{PH}}=\bigoplus_{\bj}\mathbb{I}_{2}\otimes\sigma^{x}$.

The Hamiltonian \eqref{eqn:realham} can be rewritten in the diagonal basis
\begin{equation}\label{eqn:diagham}
    H = \sum_{n}\Big(E_{n}\tilde{a}_{n}^{\da}\tilde{a}_{n}+E_{-n}\tilde{a}_{-n}^{\da}\tilde{a}_{-n}\Big),
\end{equation}
where $\tilde{a}_{n}=\bra{\kappa_{n}}\Psi$ and $\tilde{a}_{n}=\tilde{a}_{-n}^{\da}$ (see Appendix \ref{app:reptrans}). These `spectral' Dirac fermions obey the canonical fermionic anticommutation relations. We can decompose these fermions into spectral Majorana modes in the following way
\begin{equation}
    \ta_{n} = \frac{\tgam_{1,n}+i\tgam_{2,n}}{2},
\end{equation}
where $\tgam_{g,n}=\tgam_{g,n}^{\da}$ and $\{\tgam_{g,n},\tgam_{g',n'}\}=2\delta_{nn'}\delta_{gg'}$. In this basis Hamiltonian \eqref{eqn:diagham} is given by
\begin{equation}
    H = i\sum_{n}E_{n}\tgam_{1,n}\tgam_{2,n},
\end{equation}
up to a constant shift in energy. It is possible to extract the amplitude of a given spectral Majorana fermion at a given site $\bj$, hereby denoted $|\Gamma_{g,n}(\bj)|$, from the eigenstates of the kernel $\Lambda$ (see Appendix \ref{app:reptrans}).

Each time we introduce a defect the Majorana modes may become localised around it. To describe this behaviour we introduce a vector of real parameters, $\bs{\xi} = (\xi_{x},\xi_{y},\xi_{z})$, that describes the localisation of a Majorana mode around a defect point, line or surface, in the following way. A spectral Majorana mode will decay as
\begin{equation}
    |\Gamma_{g,n}(\bs{J})|\propto e^{-\|\bs{\xi}\cdot(\bj-\bj_{0})\|},
\end{equation}
where  $\xi_{x},\xi_{y},\xi_{z}\in\mathbb{R}$ are inversely proportional to the localisation length along their corresponding direction and $\|\bs{\xi}\cdot(\bj-\bj_{0})\|=\sum_{i=x,y,z}|\xi_{i}(j_{i}-(j_{0})_{i})|$. The parameter $\bj_{0}$ defines  the point, line or plane to which a state is localised. For example, a state localised to the plane located at $z=N_{z}$ has $\bj_{0}=j_{x}\bxhat+j_{y}\byhat+N_{z}\bzhat$, where $N_{z}\bzhat$ define the position of the plane in the $z$ direction and $j_x$, $j_y$ are the coordinates on the plane. On the other hand a state localised in the $x$ direction to a line that passes through the point $(0,a,b)$ has $\bj_{0}=j_{x}\bxhat+a\byhat+b\bzhat$. 

\section{2D Majorana Modes}\label{sec:2D}

\subsection{Open boundaries}

\begin{figure}[t]
\begin{center}
    \includegraphics[scale=1.0]{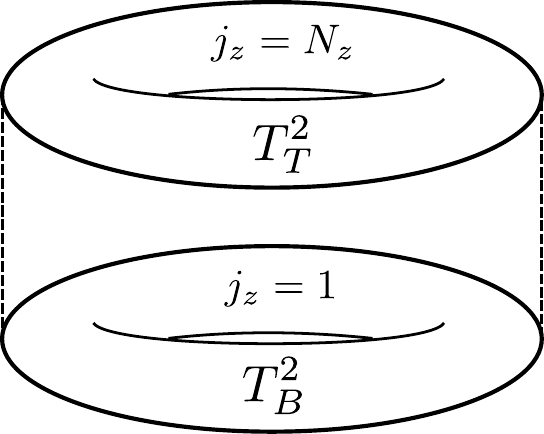}
\end{center}
\caption{A schematic representation of the boundary of the 3D superconductor. Due to the periodic boundary conditions in the $x$ and $y$ directions, the top and bottom surfaces that constitute the boundary can be viewed as a pair of tori, $T^{2}_{T}$ located at $j_{z}=N_{z}$, and $T_{B}^{2}$ located at $j_{z}=1$.}
\label{fig:TORUSSURF}
\end{figure}

We initialise the system in the $\nu_{\text{3D}} = 1$ phase ($\mu=3$, $t=\Delta=1$) and break the periodic boundary condition in the $z$ direction. The system's boundary can be viewed as a pair of toroidal surfaces $T^{2}_{T}$, located at $j_{z}=N_{z}$, and $T^{2}_{B}$, located at $j_{z}=1$, as shown in Fig.~\ref{fig:TORUSSURF}. The energy spectrum of the kernel $\Lambda$ acquires a set of four mid-gap states $\{\ket{\kappa_{1}},\ket{\kappa_{2}},\ket{\kappa_{-1}},\ket{\kappa_{-2}}\}$ that have eigenvalues $\{E_{1},E_{2},-E_{1},-E_{2}\}$ respectively. In the current configuration we have $E_{1}=E_{2}=0$. The four states correspond to two spectral Dirac fermions $\ta_{i}$, where $i=1,2$, and their PH symmetric partners. As previously asserted, these spectral Dirac fermions can be decomposed into four spectral Majorana modes $\tgam_{g,1}$ and $\tgam_{g,2}$. If we plot the spatial distribution of these four gapless Majorana modes, we find that $\tgam_{1,1}$ and $\tgam_{1,2}$ are supported on $T^{2}_{T}$, while $\tgam_{2,1}$ and $\tgam_{2,2}$ are supported on $T^{2}_{B}$. All of the Majoranas are completely delocalised in the $x$-$y$ plane while being exponentially localised to their respective surfaces in the $z$ direction. This is succinctly expressed as
\begin{equation}
    |\Gamma_{1,n}(\bj)|\propto e^{-\|\bs{\xi}\cdot(\bj-\bj_{0,T})\|}, \quad |\Gamma_{2,n}(\bj)|\propto e^{-\|\bs{\xi}\cdot(\bj-\bj_{0,B})\|},
\end{equation}
where $\bj_{0,T}=j_{x}\bxhat+j_{y}\byhat+\bzhat$, $\bj_{0,B}=j_{x}\bxhat+j_{y}\byhat+N_{z}\bzhat$, and $\bs{\xi}=(0,0,\xi_{z})$. Numerical evaluation finds that $\xi_{z}\approx1.6$. 

\subsection{Effective Zeeman field}

We can induce an energy splitting in the gapless states by introducing an effective Zeeman field at the boundary. The Hamiltonian \eqref{eqn:realham} becomes  
\begin{equation}
    H' = H + H_{\Omega}, \quad H_{\Omega}=\sum_{\bj}\bs{\psi}^{\da}\bs{B}\cdot\bs{\sigma}\bs{\psi},
\end{equation}
where $\bs{\psi}=\bpm a_{1,\bj} & a_{2,\bj} \epm^{T}$ and $\boldsymbol{B}=(B_{x},B_{y},B_{z})$ where $\boldsymbol{B}=0$ for $1<j_{z}<N_{z}$. We find that $\Delta E \propto B_{y}$  for $0\leq B_{y}< 1.5$. The other parameters $B_{x}$ and $B_{z}$ change the position of the Dirac cones in momentum space \cite{Finch15}, but have no effect on the magnitude of $\Delta E$. Due to the bulk-boundary correspondence \cite{Teo10b} the edge states are separated from the bulk states and only a phase transition can mix them. As a consequence, we can consider how the boundary Hamiltonian $H_\Omega$ acts on the edge states alone. We can achieve that by projecting the boundary Hamiltonian onto the surface states $\ket{\psi_{i}}=2\tilde{a}_{i}^{\da}\bigotimes_{\bj}\ket{0_{a_{1}}0_{a_{2}}}$, where $\bigotimes_{\bj}\ket{0_{a_{1}}0_{a_{2}}}$ is the fermionic vacuum. The resulting effective Hamiltonian is given by
\begin{equation}
    H_{\text{eff}}=\mathcal{N}\sum_{ij}\braket{\psi_{i}|H_{\Omega}|\psi_{j}}\ket{\psi_{i}}\bra{\psi_{j}},
\end{equation}
where $\mathcal{N}$ is a fitting parameter that is dependent on system size. We introduce it to account for the fact that the effective Hamiltonian is a two-dimensional object derived from three-dimensional states. The eigenvalues of $H_{\text{eff}}$, denoted $\Delta\tilde{E}_{F}$, are degenerate and numerical evaluation finds that they are equal to the energy splitting $\Delta E$. The Zeeman field induces a position dependent, local coupling between the spatially varying (pseudo-)spin degrees of freedom. One can rewrite the $H_{\text{eff}}$ in terms of Majorana operators and show that the effective Zeeman field couples all the various gapless Majorana modes. Given that the effective Zeeman field term is local, only zero energy Majoranas that have support on the same site will contribute to the energy splitting. Further to this, if the spatial distribution of the surface states $\ket{\psi_{i}}$ remains constant over the $x$-$y$ plane as a function of $B_{y}$ then $\Delta \tilde{E}\propto B_{y}$. This is consistent with our previous findings \cite{Finch15}.
\begin{figure}[t!]
\begin{center}
    \includegraphics[scale=1.0]{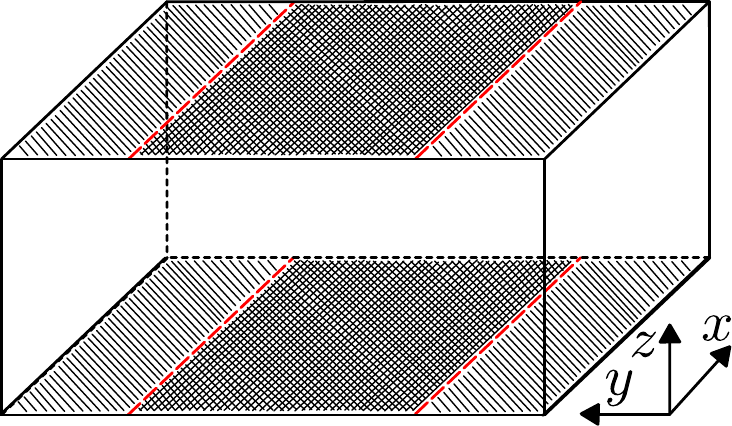}\\
    \vspace{0.5cm}
    \includegraphics[scale=1.0]{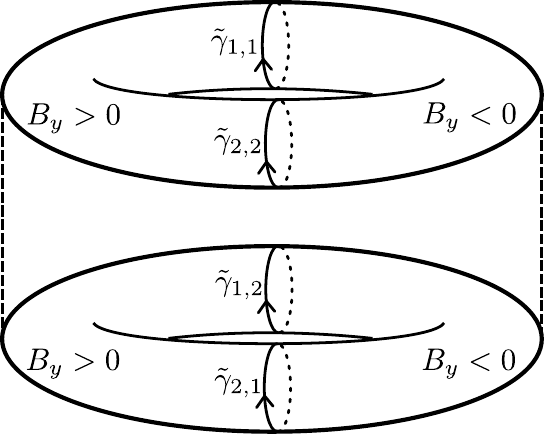}
\end{center}
\caption{(Top) A schematic representation of the configuration of local effective Zeeman fields when localising the Majorana modes to 1D. The system is periodic in the $x$ and $y$ directions while having open boundaries in the $z$ direction. A locally varying effective Zeeman field has been applied such that $\text{sgn}(B_{y})=1$ for $N_{y}/4\leq j_{y} \leq 3N_{y}/4$ and $\text{sgn}(B_{y})=-1$ elsewhere. Gapless Majorana modes appear at the interfaces between the different local effective Zeeman fields, indicated by the red dashed lines. (Bottom) The four gapless Majorana modes localised to the four interfaces between regions of differing effective Zeeman field.}
\label{fig:1DScheme}
\end{figure}

\begin{figure}[t!]
\begin{center}
    \includegraphics[scale=0.45]{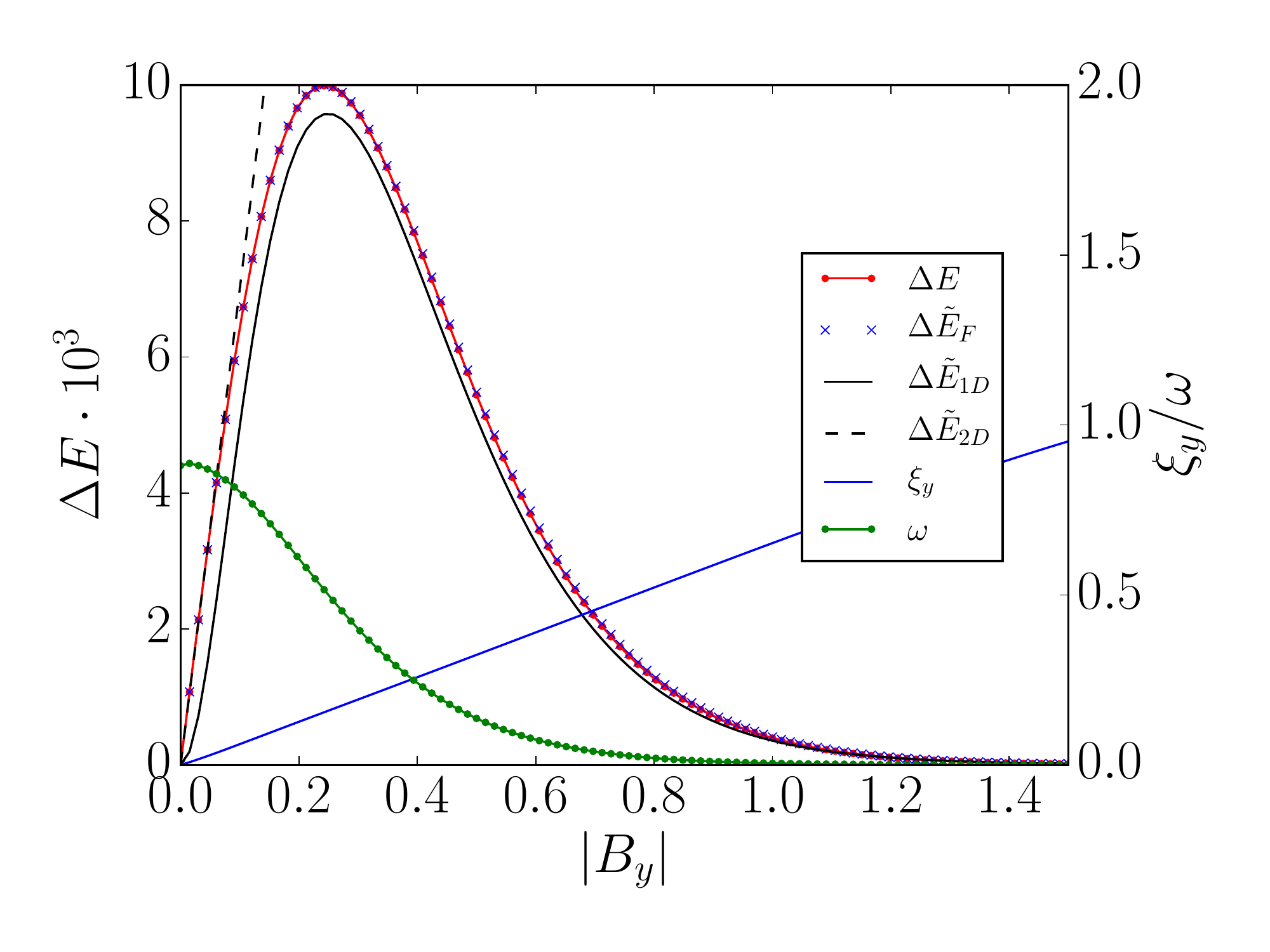}\\
    \includegraphics[scale=.45]{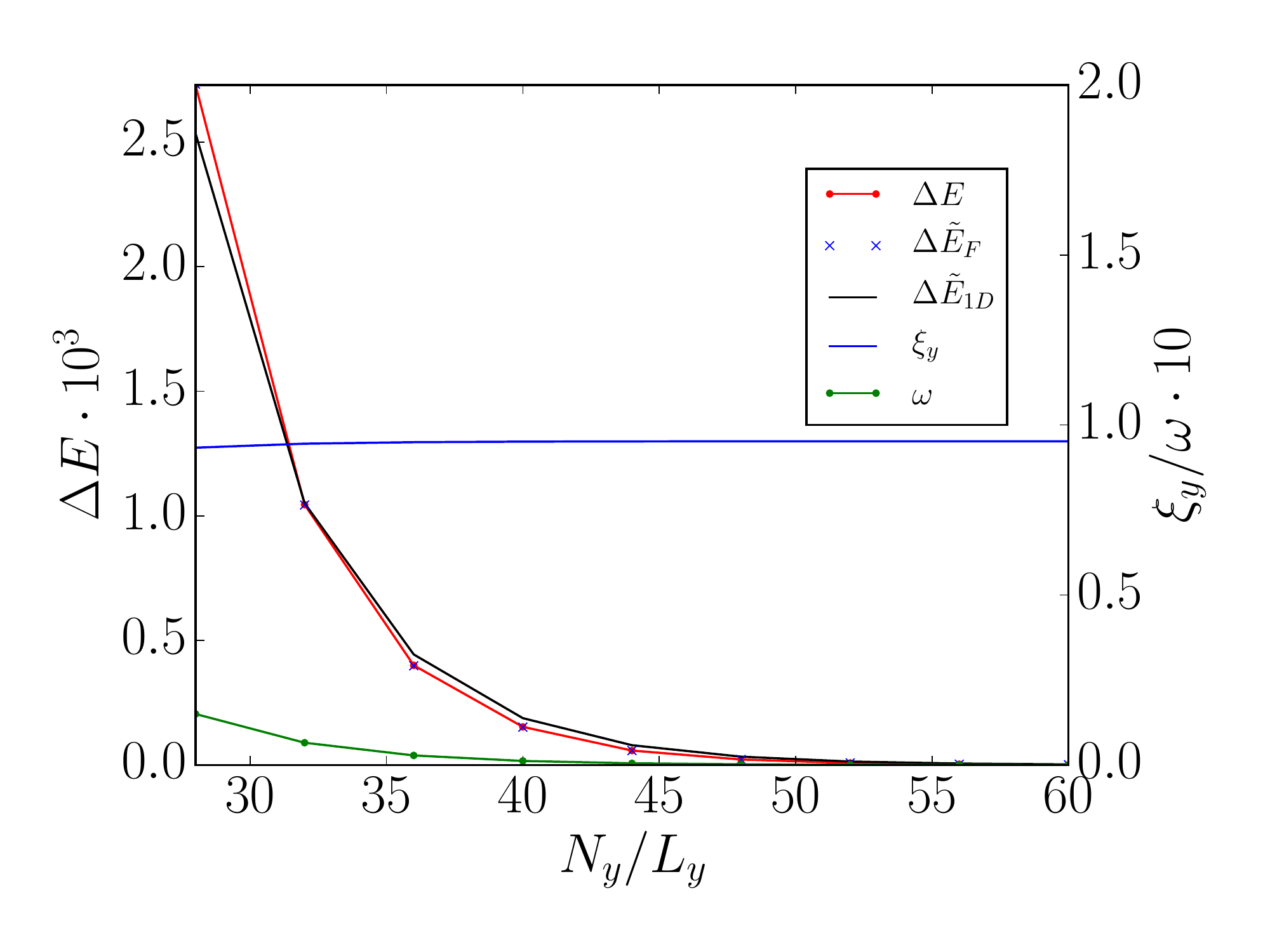}
\end{center}
\caption{(Top) A plot of the energy splitting $\Delta E$ and the localisation in the $y$ direction $\xi_{y}$ as a function of $|B_{y}|$ ($25\times50\times14$, $\mu=3$, $\Delta=t=1$) for the 1D defect configuration. The quantity $\omega=\Delta\tilde{E}_{F}/B_{y}$  decreases exponentially as the states are progressively localised and their spatial overlap decreases. The eigenvalue of $H_{\text{eff}}$ is shown as $\Delta\tilde{E}_{F}$ and corresponds exactly to $\Delta E$. $\Delta\tilde{E}_\text{1D}$ and $\Delta\tilde{E}_\text{2D}$ correspond to the theoretical predictions for the energy splitting based on the ansatz wave functions $\psi^\text{1D}_{i}(\bs{r})$ \eqref{eqn:ansatz1D} and $\psi^\text{2D}_{i}(\bs{r})$ \eqref{eqn:flat}. (Bottom) The energy gap $\Delta E$ and coherence length $\xi_{y}$ for the system in the 1D defect configuration while varying the system size in the $y$ direction, for $|B_{y}|=1.5$. As the system size increases the size of the $\text{sgn}(B_{y})$ region remains $N_{y}/4 <j_{y}<3N_{y}/4$. As the overlap between the gapless modes decreases the energy gap, $\Delta E$, decreases exponentially. The predicted energy gap, $\Delta\tilde{E}$, accurately reproduces the numerically observed values.}
\label{fig:1DXi}
\end{figure}

\section{1D Majorana Zero Modes}\label{sec:1D}

\subsection{Effective Zeeman field configuration}

We now present a scheme for reducing the dimensionality of the gapless Majorana modes from two to one dimension. We modify the effective Zeeman field such that $\text{sgn}(B_{y})=1$ for $N_{y}/4 <j_{y}<3N_{y}/4$ and $\text{sgn}(B_{y})=-1$ elsewhere. This creates a pair of magnetic domain walls on each surface of the boundary. This configuration is depicted in Fig.~\ref{fig:1DScheme}~(Top). If we initialise the system with a large magnitude Zeeman field,  we find that the four mid-gap states are still present. The spatial distribution of the four gapless Majorana modes has changed. Each of the four Majoranas are still localised to the same tori as in the uniform field configuration. However, the modes on each surface are now spatially separated such that each is exponentially localised to a different single magnetic domain wall, as depicted in Fig.~\ref{fig:1DScheme} (Bottom). In terms of the previously defined notation the Majorana modes are localised as
\begin{align}
    |\Gamma_{1,1}(\bj)| \propto e^{-\|\bs{\xi}\cdot(\bj-\bj_{0,\text{T},1})\|}, \quad |\Gamma_{2,1}(\bj)| \propto e^{-\|\bs{\xi}\cdot(\bj-\bj_{0,\text{B},2})\|},\nn\\
    |\Gamma_{1,2}(\bj)| \propto e^{-\|\bs{\xi}\cdot(\bj-\bj_{0,\text{T},2})\|}, \quad |\Gamma_{2,2}(\bj)| \propto e^{-\|\bs{\xi}\cdot(\bj-\bj_{0,\text{B},1})\|},
\end{align}
where $\bj_{0,\text{T},1}=j_{x}\bxhat+\frac{N_{y}}{4}\byhat+N_{z}\bzhat$, $\bj_{0,\text{T},2}=j_{x}\bxhat+\frac{3N_{y}}{4}\byhat+N_{z}\bzhat$, $\bj_{0,\text{B},1}=j_{x}\bxhat+\frac{N_{y}}{4}\byhat+\bzhat$ and $\bj_{0,\text{B},2}=j_{x}\bxhat+\frac{3N_{y}}{4}\byhat+\bzhat$ and coherence lengths $\bs{\xi}= (0,\xi_{y},\xi_{z})$, with $\xi_{z}\approx1.90$ and $\xi_{y}\approx0.94$.

\begin{figure}[t]
\begin{center}
    \includegraphics[scale=0.4]{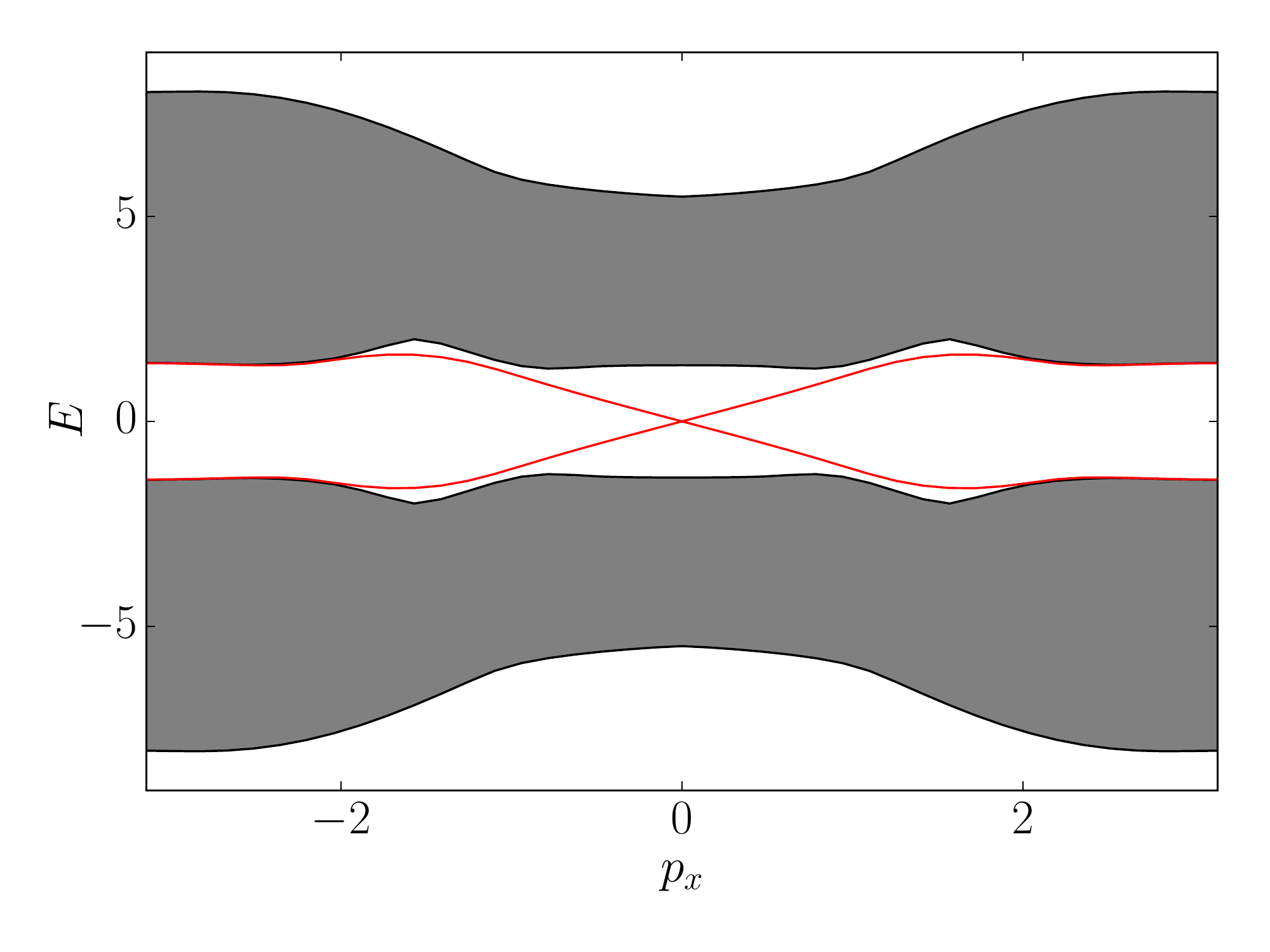}
\end{center}
\caption{The energy dispersion of the system in the 1D defect configuration ($N_{y}=50$, $N_{z}=14$, $\mu=3$, $\Delta=t=1$), having Fourier transformed the system in the $x$ direction. The system supports two pairs of degenerate counter propagating gapless Majorana modes indicated in red, each localised at a different defect line. The grey regions contain the bulk bands, not explicitly shown here.}
\label{fig:momspec}
\end{figure}

Fig. \ref{fig:1DXi}~(Top) depicts $\xi_{y}$ and $\Delta E$ (computed via exact diagonalisation) as a function of $|B_{y}|$. The localisation parameter $\xi_{y}$ was computed directly from the spatial amplitudes $|\Gamma_{g,n}(\bj)|$ and was found to be proportional to $|B_{y}|$. The energy splitting follows a linear increase which transitions into an exponential decay to zero. The eigenvalue of the effective Hamiltonian $H_{\text{eff}}$ is also shown in Fig.~\ref{fig:1DXi}; it exactly corresponds to the energy splitting as computed via exact diagonalisation. The behaviour of the energy splitting can be attributed to the gapless Majorana modes on each surface coupling in the presence of the effective Zeeman field, while simultaneously being localised by that same field. This is made plain if we plot $\omega=\Delta\tilde{E}_{F}/B_{y}$, as shown in Fig.~\ref{fig:1DXi}. As the magnitude of the effective Zeeman field increases, the spatial overlap in the presence of the coupling field between the different states decreases exponentially.

\subsection{Theoretical model}

We can further reinforce this picture by theoretically modelling this behaviour. We define four ansatz wave functions $\psi_{i}(\bs{r})$ with $i=1,2,3,4$, trapped at four potentials of height $|B_{y}|$ located at $\bs{r}_{0}^{1}={L_{y}\over 4} \bs{\hat{y}}+L_{z}\bs{\hat{z}}$, $\bs{r}_{0}^{2}={3L_{y}\over 4}\bs{\hat{y}}+L_{z} \bs{\hat{z}}$, $\bs{r}_{0}^{3}={3L_{y}\over 4} \bs{\hat{y}}+\bs{\hat{z}}$, and $\bs{r}_{0}^{4}={3L_{y}\over 4} \bs{\hat{y}}+\bs{\hat{z}}$, where $L_{x}$, $L_{y}$ and $L_{z}$ are the system sizes in the $x$, $y$ and $z$ directions. The states can be written as
\begin{equation}\label{eqn:ansatz1D}
    \psi_{i}^\text{1D}(\bs{r})=\sqrt{\frac{\xi_{y}\xi_{z}}{2L_{x}\bar{Y}\bar{Z}}}e^{-\frac{\xi_{y}}{2}|y-y_{i}|}e^{-\frac{\xi_{z}}{2}|z-z_{i}|},
\end{equation}
where $\bar{Y}=(1-e^{-\xi_{y}L_{y}})$ and $\bar{Z}=(1-e^{-\xi_{z}L_{z}})$. If we calculate the energy shift associated with the overlap between the two states on each surface, neglecting overlaps between states on different $z$ levels, we have
\begin{equation}
    \Delta\tilde{E}_\text{1D}=4\sum_{i\neq j}\int_{0}^{L_{x}}\int_{\frac{L_{y}}{4}}^{\frac{3L_{y}}{4}}\int^{l}_{0}dz\,dy\,dx\,\,\,\,\psi_{i}^\text{1D}(\bs{r})V_{0}\psi_{j}^\text{1D}(\bs{r}),
\end{equation}
where we have periodic boundary conditions and $V_{0}=|B_{y}|(1-\text{H}(l-z))$, where $\text{H}(l-z)=\int_{-\infty}^{l-z}\delta(s)ds$ is the Heaviside function and $|B_{y}|$ is the strength of the potential that couples the 1D localised states. The Heaviside function is introduced as the Zeeman field is applied exclusively on the surface. The parameter $l$ reflects the fact that in the continuum case the Zeeman field penetrates the bulk. We use $l$ as a fitting parameter to map the continuum model to the discrete case. Evaluating this we find that
\begin{equation}
    \Delta\tilde{E}_{\text{1D}}=\xi_{y}|B_{y}|L_{y}e^{-\frac{\xi_{y}}{4}L_{y}}\big(1-e^{-\xi_{z}l}\big).
\end{equation}

\noi The plot of $\Delta\tilde{E}_\text{1D}$ as a function of $|B_{y}|$ is shown in Fig.~\ref{fig:1DXi}~(Top).  By a numerical fit to the exact diagonalisation data we find that $l\approx L_{y}^{-1}$. In the region $|B_{y}|>0.2$ of Fig.~\ref{fig:1DXi}, there is a strong agreement between the behaviour of the numerical observations and the modelling of the states by \rf{eqn:ansatz1D}. We note that the value of $\xi_{y}$ is derived from the numerical data of the state amplitudes only on the boundary. Similarly, the value of $\xi_{z}$ is taken to be an average over the surface.  A more careful analysis would require $\xi_{y}=\xi_{y}(z)$ and $\xi_{z}=\xi_{z}(y)$, which might explain the slight deviation between the numerical and the predicted values of $\Delta E$. By averaging over the relevant spatial variables we find a good match for the overall behaviour of the surface physics. We also apply this formula for varying $L_{y}$, as shown in Fig.~\ref{fig:1DXi}~(Bottom). In the region $|B_{y}|<0.2$, the states are transitioning between their extended 2D form and an exponentially localised 1D form.  For small values of $|B_{y}|$, the states are closer to the form
\begin{equation}\label{eqn:flat}
    \psi_{i}^\text{2D}(\bs{r}) = \sqrt{\frac{\xi_{z}}{L_{x}L_{y}\bar{Z}}}e^{-\frac{\xi_{z}}{2}|z-z_{i}|}.
\end{equation}
The overlap of two of these states on the same surface in the presence of the potential $V_{0}=|B_{y}|(1-\text{H}(l-z))$ is given by
\begin{equation}
    \Delta \tilde{E}_{\text{2D}}=|B_{y}|\big(1-e^{-\xi_{z}l}\big).
\end{equation}
The behaviour of this model is shown in Fig.~\ref{fig:1DXi}, where we have again taken $l\approx L_{y}^{-1}$. The initial linear increase in $\Delta E$ is accurately reproduced by assuming that the states are spread evenly throughout the 2D surface, as modelled by \eqref{eqn:flat}.

\subsection{1D reciprocal space}

To determine the discretion relation of the 1D Majorana modes, we Fourier transform the Hamiltonian in the $x$ direction, giving
\begin{equation}
    H'= \sum_{p_{x}}\Psi_{p_{x}}^{\da}\Lambda(p_{x})\Psi_{p_{x}},
\end{equation}
where $\Psi_{p_{x}}=\bigoplus_{\bar{\bj}}\bpm a_{1,\bar{\bj},p_{x}},&a_{1,\bar{\bj},p_{x}}^{\da},&a_{1,\bar{\bj},p_{x}},& a_{1,\bar{\bj},p_{x}}^{\da} \epm^{T} $ with $\bar{\bj}=\bpm j_{y}, &j_{z} \epm $ and $p_{x}\in[-\pi,\pi)$. The energy dispersion as a function of $p_{x}$ is shown in Fig.~\ref{fig:momspec}. The spectrum is PH symmetric and the bulk negative energy states are separated from the bulk excited states by a bulk energy gap. There is a set of eigenvalues that cross zero energy at $p_{x}=0$. Because, the mapping that takes us to the spectral Majorana modes is unitary, the crossing states correspond to the four gapless Majorana modes $\tgam_{g,n}$ now given by $\tgam_{g,n,p_{x}}$. In reciprocal space, the Majorana modes no longer obey the reality condition and we have $\tgam_{g,n,p_{x}}=\tgam_{g,n,-p_{x}}^{\da}$. This implies that the spectrum for $p_{x}>0$ is an inverted copy of $p_{x}<0$. Furthermore, the eigenvalue of $\tgam_{1,n,p_{x}}$ is equal in magnitude to the eigenvalue of $\tgam_{2,n,p_{x}}$ but with opposite sign. This leads us to conclude that $\tgam_{1,1}$ and $\tgam_{1,2}$ propagate in the same direction, but in the opposite direction to $\tgam_{2,1}$ and $\tgam_{2,2}$, as shown in Fig.~\ref{fig:1DScheme} (Bottom).

\begin{figure}
    \begin{center}
        \includegraphics[scale=.6]{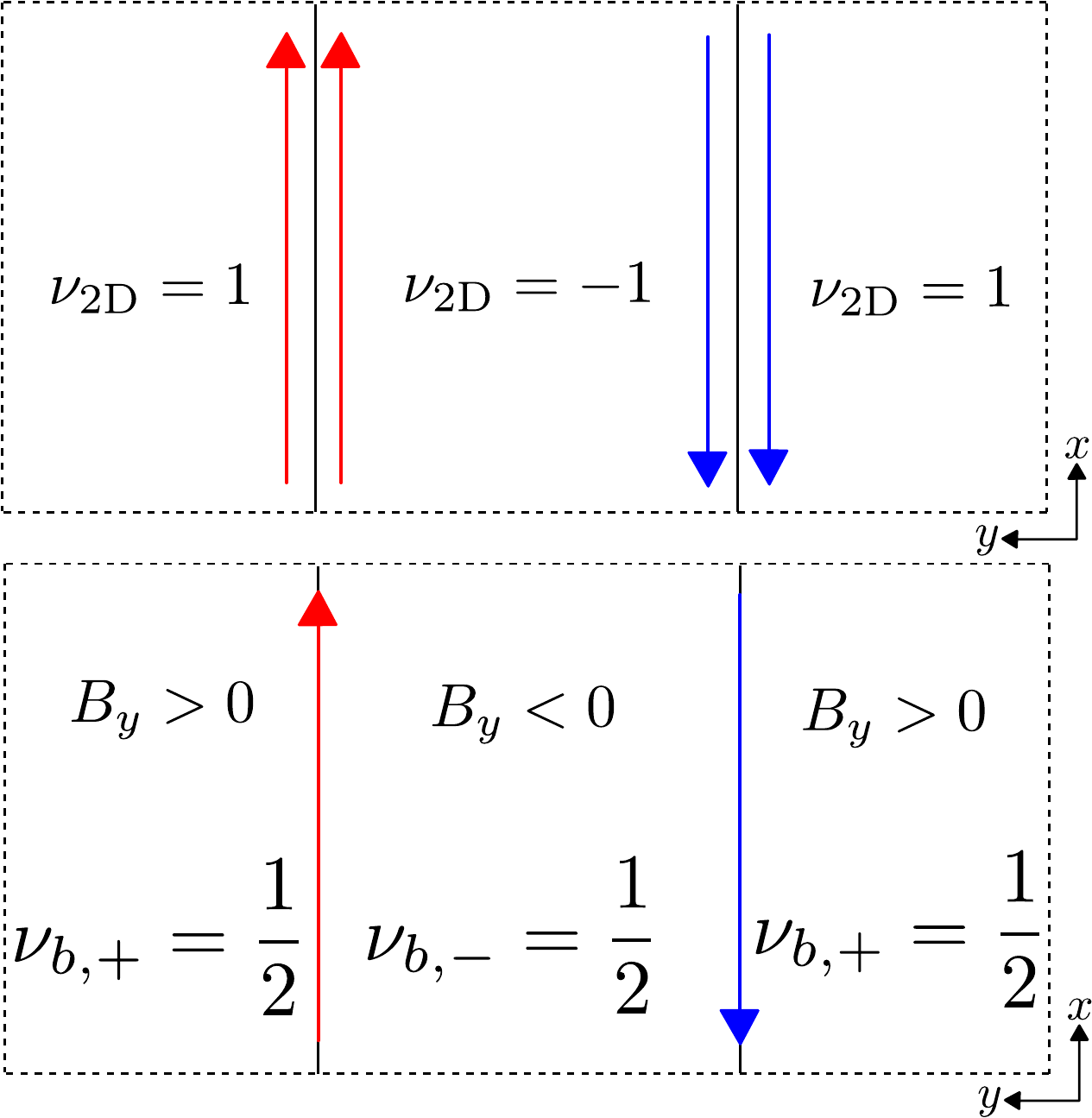}
    \end{center}
    \caption{\label{fig:2DD} (Top) A schematic representation of a pair of 2D class D TS on a torus. The dashed lines indicate a periodic boundary. The Chern numbers of the systems differ by a sign. Such chiral superconductors possess a pair of chiral counter-propagating Majorana modes at their boundaries. The orientation of the modes is defined up to a sign in the Chern number. (Bottom) A schematic representation of a single surface of our 3D DIII system. The two magnetic domain walls trap one Majorana fermion each. The number of Majorana modes is the difference of the partial Chern numbers in each magnetic domain.  }
\end{figure}

\subsection{Fractional Chern numbers and edge states}

The behaviour of the surface states in the presence of magnetic domain walls is commensurate with our understanding of chiral 2D TS. Given a pair of 2D class D topological superconductors on a torus, the number of gapless Majorana states at the interface is equal to the difference of their Chern numbers, as given in \eqref{eqn:index} \cite{Beri10}. This configuration is shown in Fig.~\ref{fig:2DD} (Top). In the presence of a uniform non-zero effective Zeeman field the union of the top and bottom surfaces of the 3D DIII model can be considered to be a 2D class D TS~\cite{Finch15}. The sign of the partial contribution to the Chern number of each boundary surface is equal to the sign of $B_{y}$ at that surface. When the magnetic domain walls are introduced we can effectively view each magnetic domain as one-half of a 2D class D superconductor in its own right. The systems either side of a domain wall have partial Chern numbers of opposite sign, as shown in Fig. \ref{fig:2DD} (Bottom). We find that the number of Majorana modes at each interface is equal to the difference in the partial Chern numbers in each magnetic domain on a surface. This suggests the relation
\begin{equation}\label{eqn:indexfrac}
    N=|\nu_{b,+}-\nu_{b,-}|,
\end{equation}
where $\nu_{b,\pm}$ are the partial Chern numbers either side of a magnetic domain wall. 

We can demonstrate the validity of \rf{eqn:indexfrac} by starting from relation \rf{eqn:index} that gives the number of edge states between surfaces with integer Chern numbers. It is known that the Chern number of a composite system of two non-interacting 2D subsystems is equal to the sum of the Chern numbers of the two subsystems. This property holds true even if one of the Chern numbers is half integer, e.g. when one of the subsystems is the 2D boundary of a 3D system. Consider now two neighbouring TS with bulk invariants $\nu_+$ and $\nu_-$ with $N=|\nu_+-\nu_-|$ gapless states at their interface. We superpose to them a gapped 2D surface of a 3D system that has Chern number $\nu_b=1/2$. This surface supports no edge modes hence it will not change the total number $N$ of edge states. But it will modify the total Chern number at either side of the interface to become half-integer, giving eventually \rf{eqn:indexfrac}.

It is also possible to show the consistency of this new relationship with the original index theorem. The Chern number of the whole boundary of a system is defined as $\nu_{\text{2D}}=\sum_{b}\nu_{b}$. Take two copies of the 3D bulk system with Zeeman field of opposite sign. One of the systems has a boundary with Chern number $\nu_{\text{2D},+}=\sum_{b}\nu_{b,+}$ while the other has a boundary with Chern number $\nu_{\text{2D},-}=\sum_{b}\nu_{b,-}$. We now glue them together such that we have periodic boundary conditions in the $x$ and $y$ directions with the 1D defect configuration on the boundary of the composite system, as described previously. The index theorem \eqref{eqn:index} relates the Chern numbers of the boundaries of both systems (the union of the top and bottom surfaces of each) to the number of states that appear at their interface. In the case when $\nu_{\text{3D}}=1$ \eqref{eqn:index} tells us there should be two states at the interface. We can rewrite \eqref{eqn:index} in the following way
\begin{align}
    N=&\,\,|\nu_{\text{2D},+}-\nu_{\text{2D},-}|\nn\\
      =&\,\,|\nu_{T,+}+\nu_{B,+}-\nu_{T,-}-\nu_{B,-}|\nn\\
      =&\,\,|\nu_{T,+}-\nu_{T,-}| + |\nu_{B,+}-\nu_{B,-}|\nn\\
      =&\,\,N_{T}+N_{B}.
\end{align}
where $N_{b}=1$. The decomposition in the second step is allowed because the quantities $\nu_{b,+}-\nu_{b,-}$ have the same sign. The original index theorem is preserved as the total number of states at the interface between the two boundary systems is 2.  

\subsection{The $\nu_\text{3D}=2$ Model}

To demonstrate the generality of our results, we now consider the defects at the boundary of a $\nu_\text{3D}=2$ TS. We consider the tight binding Hamiltonian
\begin{align}\label{eqn:HAM2}
    H_{2} &=\sum_{\bj} \Big\{\sum_{k=1,2}\Big[\mu a_{k,\bj}^\dagger a_{k,\bj} + \frac{1}{2} \nn\\
          &+ t a_{k,{\bj}}^{\da}a_{k,\bj-\bxhat} + ta_{k,\bj}^\dagger a_{k,\bj-\bxhat+\bzhat}\nn\\
          & -\Delta a_{k,{\bj}}a_{k,\bj+\bxhat}+\Delta a_{k,\bj} a_{k,\bj-\bxhat+\bzhat}\Big]\nn\\
          &-2i \Delta \big(a_{1,\bj} a_{2,\bj+\byhat}+a_{2,\bj}a_{1,\bj+\byhat}\big)\nn\\
          &+2i\Delta \big(a_{1,\bj}a_{1,\bj+\bxhat+\byhat} - a_{2,\bj}a_{2,\bj+\bxhat+\byhat} \big) + \text{H.c.} \Big\}.
\end{align}
When in the $\nu_\text{3D}=2$ phase ($\mu=\Delta=t=1$) and with open boundary conditions in the $z$ directions, the model supports two helical Majorana cones per surface. When gapped by a uniform positive effective Zeeman field, each surface contributes $\nu_{\text{T}}=\nu_{\text{B}}=1$ to the total Chern number of the boundary system such that $\nu_\text{2D}=2$.  In the language of the previous section, it supports eight mid-gap states $\{\ket{\kappa_{n}},\ket{\kappa_{-n}}\}$, $n=1,2,3,4$. This corresponds to eight gapless Majorana modes $\tgam_{g,n}$, $g=1,2$. A plot of $|\Gamma_{g,n}(\bj)|$ finds that the states are localised as
\begin{align}
    |\Gamma_{1,1}(\bj)| \propto e^{-\|\bs{\xi}\cdot(\bj-\bj_{0,\text{T},1})\|}, \quad |\Gamma_{2,1}(\bj)| \propto e^{-\|\bs{\xi}\cdot(\bj-\bj_{0,\text{B},2})\|},\nn\\
    |\Gamma_{1,2}(\bj)| \propto e^{-\|\bs{\xi}\cdot(\bj-\bj_{0,\text{T},2})\|}, \quad |\Gamma_{2,2}(\bj)| \propto e^{-\|\bs{\xi}\cdot(\bj-\bj_{0,\text{B},1})\|},\nn\\
    |\Gamma_{1,3}(\bj)| \propto e^{-\|\bs{\xi}\cdot(\bj-\bj_{0,\text{T},1})\|}, \quad |\Gamma_{2,3}(\bj)| \propto e^{-\|\bs{\xi}\cdot(\bj-\bj_{0,\text{B},2})\|},\nn\\
    |\Gamma_{1,4}(\bj)| \propto e^{-\|\bs{\xi}\cdot(\bj-\bj_{0,\text{T},2})\|}, \quad |\Gamma_{2,4}(\bj)| \propto e^{-\|\bs{\xi}\cdot(\bj-\bj_{0,\text{B},1})\|},
\end{align}
where $\bs{\xi}= (0,\xi_{y},\xi_{z})$, with $\xi_{z}=1.90$ and $\xi_{y}=0.94$. Each magnetic domain wall supports two gapless Majorana modes. The difference in the partial Chern numbers either side of a magnetic domain wall is equal to the number of states localised to it; this is commensurate with \eqref{eqn:indexfrac}.

\begin{figure}[t!]
    \begin{center}
        \includegraphics[scale=1.0]{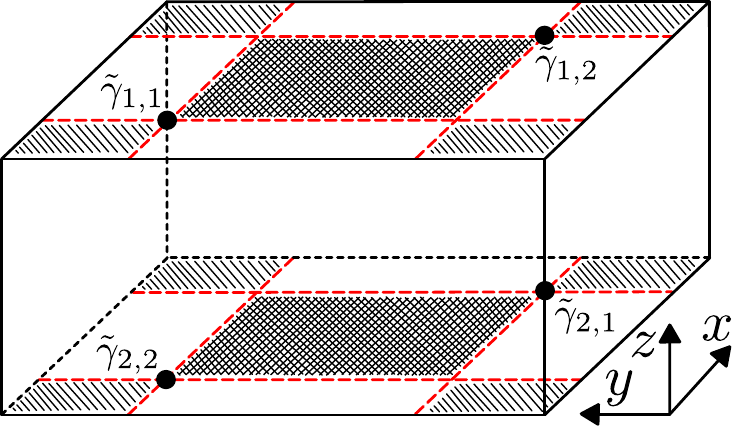}
    \end{center}
    \caption{\label{fig:0DDIAG} A schematic representation of the configuration of local effective Zeeman fields when localising the Majorana modes to 0D. The system is periodic in the $x$ and $y$ directions while having open boundaries in the $z$ direction. The black dots indicate the points at which the Majorana modes are localised. In the central darker regions $\text{sgn}(B_{y})=-1$, the lighter corner regions have $\text{sgn}(B_{y})=1$ and the white regions have $B_{y}=0$.} 
\label{fig:0DScheme}
\end{figure}

\section{Quasi-0D Majorana Modes}\label{sec:0D}

\begin{figure}[t!]
    \begin{center}
        \includegraphics[scale=0.4]{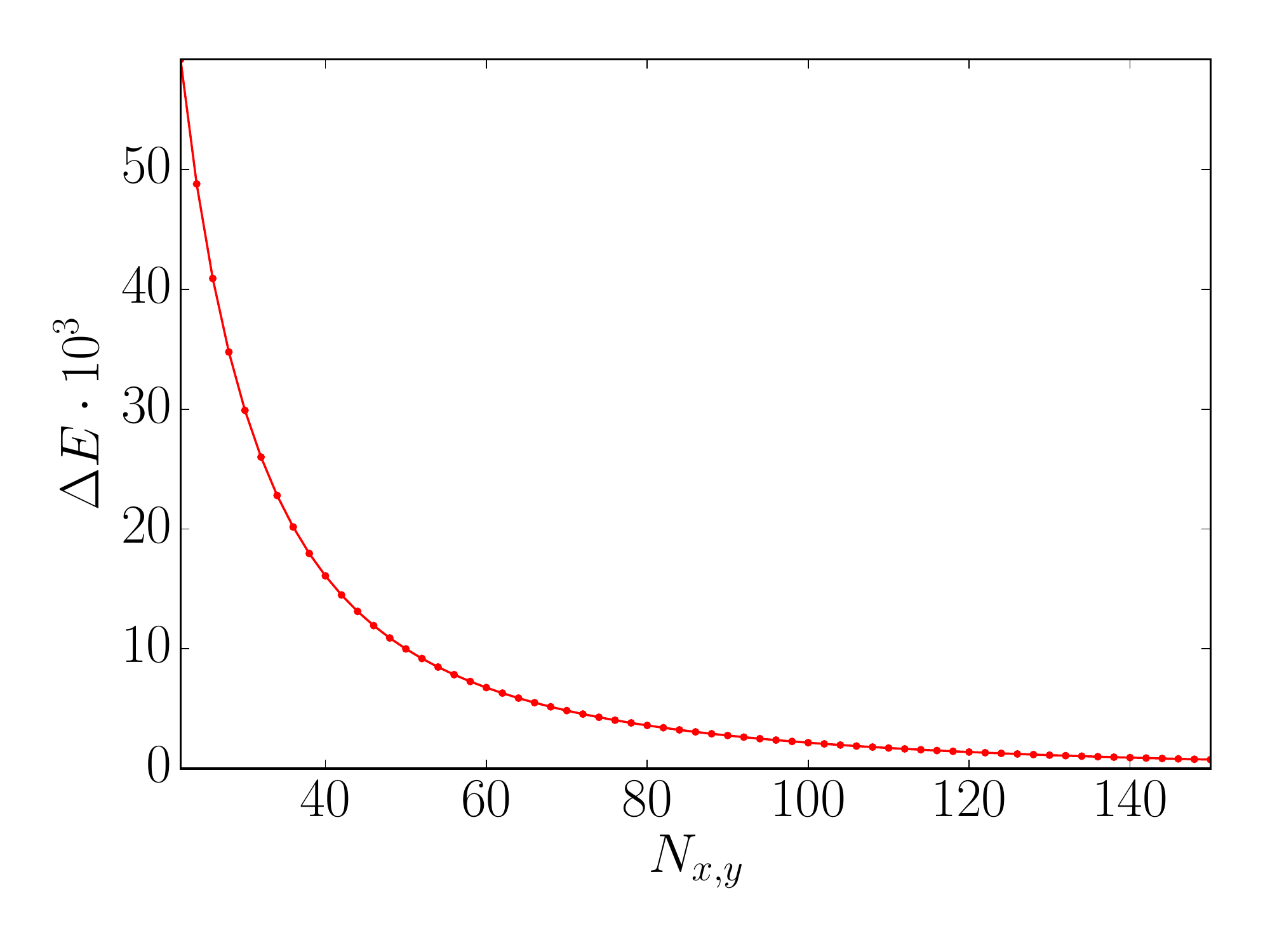}\\
        \includegraphics[scale=0.4]{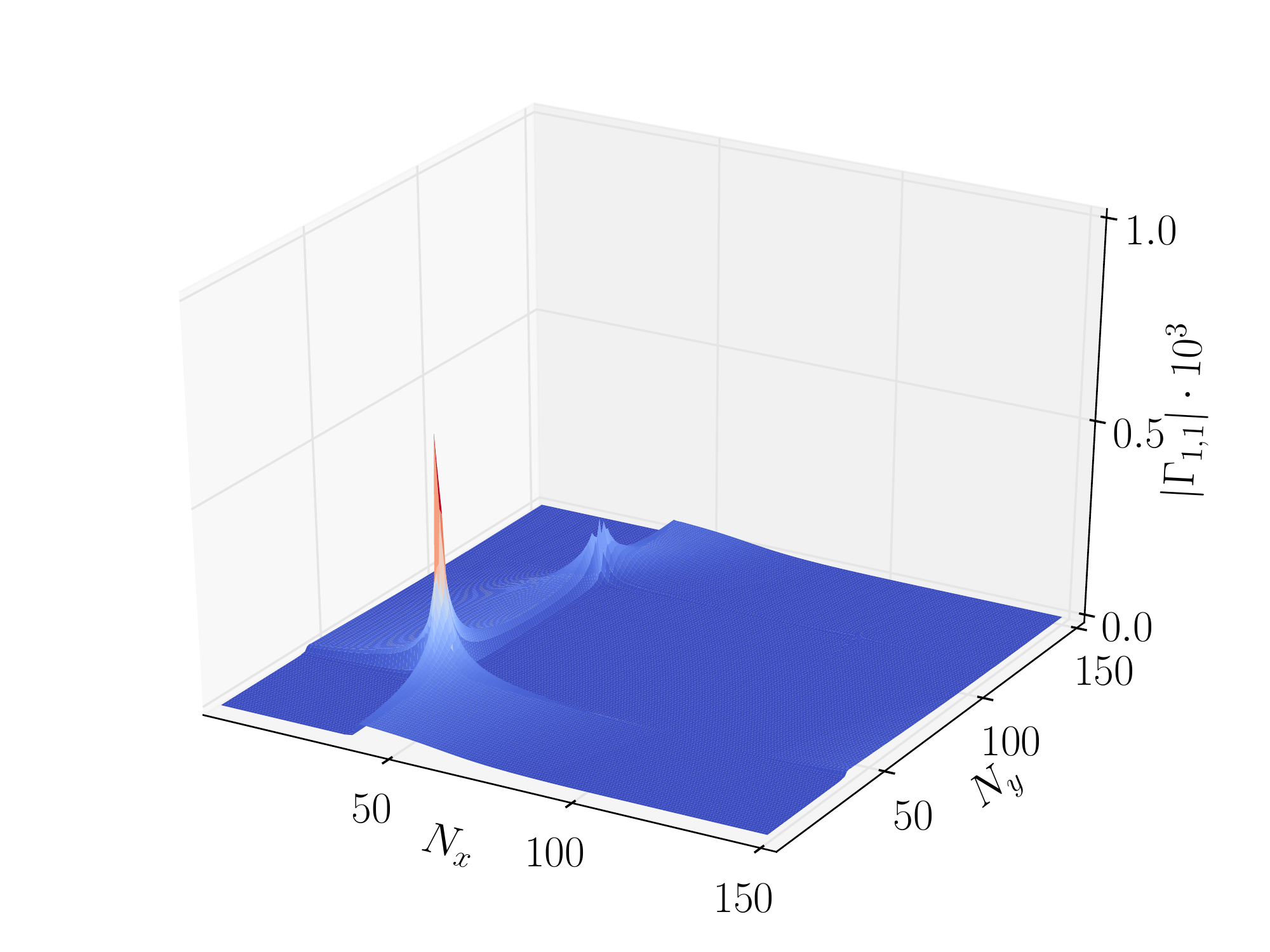}
    \end{center}
    \caption{\label{fig:0Dvaryxy} (Top) The energy splitting $\Delta E$ for chessboard configuration of boundary fields for $|B_{y}|=1.5$ for varying system size in the $x$ and $y$ directions, such that $N_{x}=N_{y}$. A numerical fit of the curve finds that $\Delta E\propto N_{x,y}^{-2}$ and will decay to zero in the thermodynamic limit. (Bottom) The amplitude of $\tgam_{1,1}$ over the $x$-$y$ plane at $z=N_{z}$.}
\end{figure}

We now consider the next iteration of nested defects where we reduce their dimensionality from 1D to 0D and study the arrangement of the surface Majorana modes in this new configuration. Taking the $\nu_{\text{3D}}=0,\pm1$ model with Hamiltonian \eqref{eqn:realham}, to produce the defects we divide the boundary surfaces into nine domains where the magnitude of the effective Zeeman field is either zero or $\pm B_{y}$, as shown in Fig.~\ref{fig:0DDIAG}. Numerical evaluation finds that the four mid-gap states $\{\ket{\kappa_{n}},\ket{\kappa_{-n}}\}$, $n=1,2$, are still present. Calculation of the amplitudes of the four associated gapless Majorana modes finds that they are localised at the following points
\begin{align}\label{eqn:0Dpoints}
\tgam_{1,1}: \quad &\bj_{0,T,a}=\frac{N_{x}}{4}\bxhat+\frac{N_{y}}{4}\byhat+N_{z}\bzhat,\nn\\
\tgam_{1,2}: \quad &\bj_{0,T,b}=\frac{3N_{x}}{4}\bxhat+\frac{3N_{y}}{4}\byhat+N_{z}\bzhat,\nn\\
\tgam_{2,1}: \quad &\bj_{0,B,a}=\frac{N_{x}}{4}\bxhat+\frac{N_{y}}{4}\byhat+\bzhat,\nn\\
\tgam_{2,2}: \quad &\bj_{0,B,b}=\frac{3N_{x}}{4}\bxhat+\frac{3N_{y}}{4}\byhat+\bzhat.
\end{align}
We expect that the Majorana modes should be exponentially localised to their respective points in regions of non-zero effective Zeeman field, as in the 1D case. In the regions of zero effective Zeeman field there is no suppressing field and as such we expect a less stringent form of localisation. Numerical evaluation of the amplitudes $|\Gamma_{g,n}|$ finds that they are not rotationally symmetric about their respective points. As expected, they are exponentially localised in the regions of non-zero effective Zeeman field and polynomially localised in the regions with zero effective Zeeman field. Due to the polynomial nature of the localisation into the regions of zero effective Zeeman fields, we say that the Majorana modes are quasi-0D. A plot of $|\Gamma_{1,1}|$ is shown in Fig. \ref{fig:0Dvaryxy}.

\begin{figure}[t!]
    \begin{center}
    \includegraphics[scale=.4]{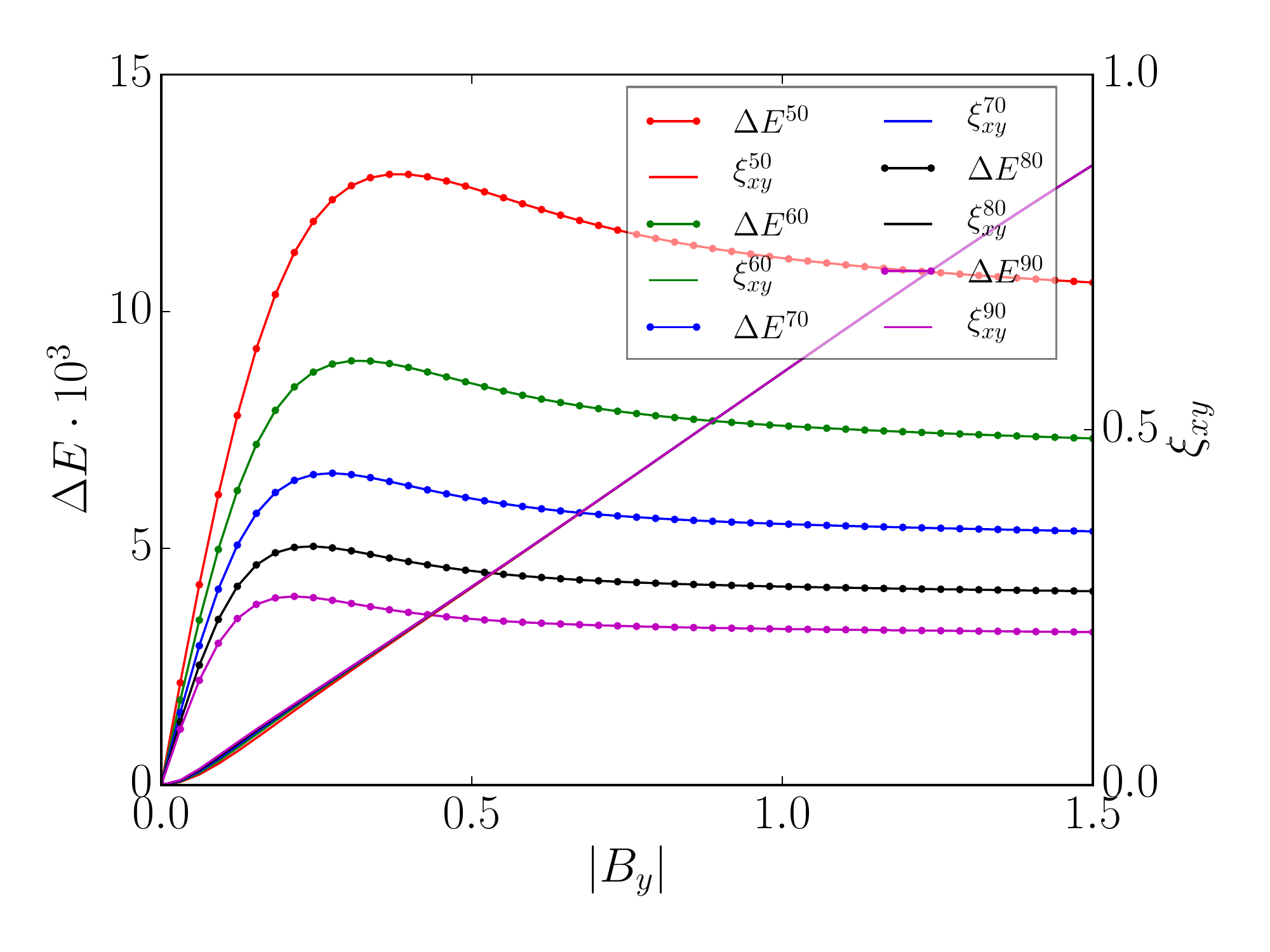}\\
    \includegraphics[scale=.4]{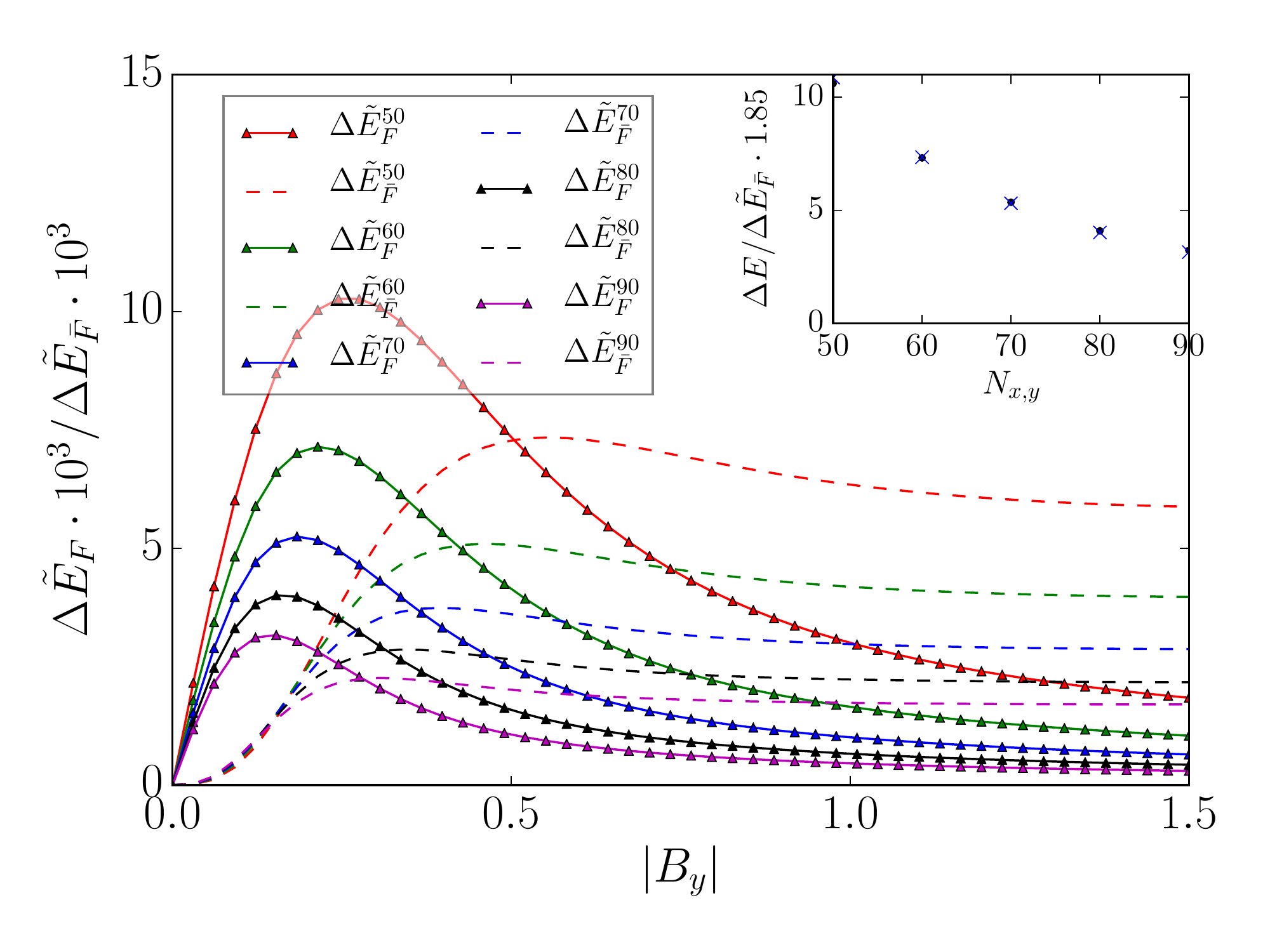}\\
    \end{center}
    \caption{\label{fig:0DVARY} (Top) The energy splitting $\Delta E$ and localisation parameter $\xi_{xy}$ as a function of $|B_{y}|$, for the chessboard configuration of effective Zeeman fields. Shown is data from systems of sizes $N_{x}=N_{y}=50$, $N_{x}=N_{y}=60$, $N_{x}=N_{y}=70$, $N_{x}=N_{y}=80$, and $N_{x}=N_{y}=90$. (Bottom) The eigenvalues of the effective Hamiltonian $H_{\text{eff}}$, $\Delta\tilde{E}_{F}$, and the corrective effective Hamiltonian $\bar{H}_{\text{eff}}$, $\Delta\tilde{E}_{\bar{F}}$. (Inset) The values of $\Delta E$ (black dots) and $\Delta\tilde{E}_{\bar{F}}$ (blue crosses) for $|B_{y}|=1.5$ for different system sizes. When $\Delta\tilde{E}_{\bar{F}}$ is multiplied by a factor of 1.85 the two quantities are the same at all system sizes.}
\end{figure}

Assuming that $N_{x}=N_{y}$, we introduce the localisation parameter $\xi_{xy}$ that refers to the exponential decay coefficient corresponding to the decay of the gapless Majorana modes along the straight line that passes through both quasi-0D Majoranas on a surface, which are shown in Fig.~\ref{fig:0DDIAG}. Fig.~\ref{fig:0DVARY} (Top) shows the energy gap $\Delta E$ and $\xi_{xy}$ for varying $|B_{y}|$. We have presented data from systems of various sizes. As $|B_{y}|$ increases, we identify in Fig.~\ref{fig:0DVARY} (Top) two distinct behaviours. For small $|B_{y}|$ we see $\Delta E$ increasing in a similar manner to the 1D case studied in Sec. \ref{sec:1D}. The states are transitioning between their 2D delocalised form to their quasi-0D form. This initial linear increase is commensurate with the gap induced by the effective Zeeman field on some 2D surface states, as previously discussed. For larger $|B_{y}|$ the energy gap exponentially decays to some constant value that depends on the system size. Fig.~\ref{fig:0Dvaryxy} (Top) shows the value of $\Delta E$ for varying system size, with $|B_{y}|=1.5$. A numerical fit finds that $\Delta E \propto N_{xy}^{-2}$, and so for sufficiently large system sizes $\Delta E\rightarrow 0$. Using this behaviour to extrapolate Fig.~\ref{fig:0DVARY}, we deduce that for a sufficiently large system size and sufficiently large effective Zeeman field we have quasi-0D \emph{zero energy} states bound at the point-like defects. 

An effective description of the surface system is two fold. The parts of the surface that experience a non-zero Zeeman field are described by the effective Hamiltonian $H_{\text{eff}}$, as in the 1D case. Its eigenvalues $\Delta\tilde{E}_{F}$ as a function of $|B_{y}|$ are shown in Fig. \ref{fig:0DVARY} (Bottom). For small values of $|B_{y}|$, $H_{\text{eff}}$ accurately reproduces the numerically evaluated energy gap $\Delta E$. For larger values of $|B_{y}|$, $\Delta\tilde{E}_{F}$ diverges from $\Delta E$ decaying too quickly as $|B_{y}|$ increases. As the magnitude of the Zeeman field becomes large the regions of the surface that experience no Zeeman field still support a significant proportion of the state. Despite the fact that they do not experience the coupling of their local degrees of freedom via the Zeeman field they still provide a contribution the energy gap. This contribution arises due to the surface states being significantly modified by the nested defects so that they are no longer zero eigenstates of the bulk Hamiltonian, $H$. This extra contribution can be seen by defining a second effective Hamiltonian  
\begin{equation}
    \bar{H}_{\text{eff}}=\sum_{\alpha\beta}\braket{\psi_{\alpha}|H|\psi_{\beta}}\ket{\psi_{\alpha}}\bra{\psi_{\beta}}.
\end{equation}
The eigenvalues of $\bar{H}_{\text{eff}}$ are degenerate and denoted $\Delta\tilde{E}_{\bar{F}}$. Fig.~\ref{fig:0DVARY} shows $\Delta\tilde{E}_{\bar{F}}$ as a function of $|B_{y}|$. For small values of $|B_{y}|$, $\Delta\tilde{E}_{\bar{F}}$ is small and does not make a significant contribution to $\Delta E$. As $|B_{y}|$ becomes large $\Delta\tilde{E}_{\bar{F}}$ converges to a constant value dependent on system size. It is this contribution which accounts for the behaviour of $\Delta E$ for large values of $|B_{y}|$. This can be seen in Fig.~\ref{fig:0DVARY} (Bottom) (Inset). The values of $\Delta E$ and $\Delta\tilde{E}_{\bar{F}}$ for $|B_{y}|=1.5$ are the same up to a constant factor of 1.85.

This field configuration traps gapless Majorana modes at 0D points that are at the interface between regions of the surface that have partial Chern numbers $\nu_{b}=\pm\frac{1}{2}$. As in the 1D case, the number of Majorana modes at each point is the difference of the partial Chern numbers either side of the interface. Where there are interfaces between regions of non-zero Zeeman field and regions that experience no Zeeman field, no well localised Majorana modes appear.  We suspect that this is because \eqref{eqn:indexfrac} is not well defined for gapless regions. An examination of this field configuration implemented in the $\nu=2$ model given in \eqref{eqn:HAM2} finds that we simply double the number of gapless Majorana modes at the four points given in \eqref{eqn:0Dpoints}.

\section{Conclusions}\label{sec:conc}

In this paper we have investigated the response of helical Majorana modes at the boundary of a 3D TS in the symmetry class DIII in the presence of nested defects. Gapless Majorana modes become localised at these defects, allowing states of dimension 2, 1 and quasi-0 to appear at the boundary of the system depending on the effective Zeeman field configuration. Furthermore, we observed a generalisation of the index theorem \eqref{eqn:index} whereby the number of gapless Majorana modes trapped at a given magnetic domain wall is the difference in the possibly partial Chern numbers either side of a wall, as defined in \eqref{eqn:FRACWIND}. This relation holds in the case of edge states localised at both 1D and 0D domain walls. 

We suggest two possible implementations of such an effective Zeeman field. First, in recent years much work has been done on implementing effective Zeeman fields through doping materials with polarised magnetic impurities \cite{Qi11,Nomura12,Foster12,Meng12,Liu12,Hu13,Deng13,Shao13,Li15,Baum15,Peng15,Cha10,Chen10,Tran10,Zitko10,Liu09,Ye10,Choy11}. Through judicious doping of sections of the surface of the superconductor, such that magnetic domain walls are created in place of the defect lines in our model, we can induce an effective Zeeman field coupling between the different pseudo-spin components of our model. Second, we could construct a superconductor to magnetic material heterostructure. By depositing a layer of ferromagnetic material on the surface of the superconductor, we can induce a spin-orbit coupling in the surface of the superconductor \cite{Mao10,Chakhalian06,Teo10b}. If a magnetic domain wall is created in the magnetic material, this would correspond to a change in the sign of the effective spin-orbit coupling, as needed of the implementation of nested defects.

\bibliographystyle{h-physrev} 
\bibliography{ref} 

\newpage

\appendix
\section{Representation Transformations}\label{app:reptrans}

\begin{figure}
    \begin{center}
        \includegraphics[scale=1]{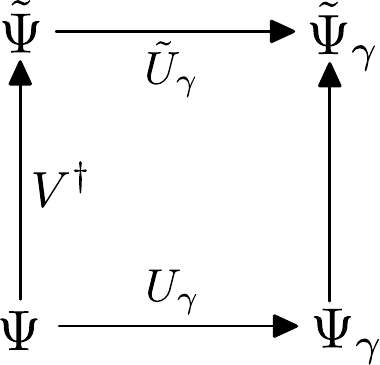}
    \end{center}
    \caption{\label{fig:TRANS}  A diagram showing the set of transformations that map between the different fermion pictures. $\Psi$ are the constituent Dirac fermions, $\Psi_{\gamma}$ are the constituent Majorana modes, $\tilde{\Psi}$ are the spectral Dirac fermions and $\Psi_{\gamma}$ are the spectral Majorana modes. $V$ is the unitary matrix found via the singular value decomposition of $\Lambda$. }
\end{figure}

There exist a set of transformations depicted in Fig.~\ref{fig:TRANS} that map between different basis representations of $H$. There are four distinct representations that are all related by a set of unitary rotations $\big\{U_{\gamma},V,\tilde{U}_{\gamma}\big\}$. The four representations are as follows
\begin{itemize}
    \item $\Psi$: constituent Dirac fermions, $a_{k,\bj}$.    
    \item $\Psi_{\gamma}$: constituent Majorana modes, $\gamma_{g,\bj}$.    
    \item $\tilde{\Psi}$: spectral Dirac fermions, $\tilde{a}_{n}$.    
    \item $\tilde{\Psi}_{\gamma}$: spectral Majorana modes, $\tilde{\gamma}_{g,n}$.    
\end{itemize}
If we start with the $H$ in the constituent fermion basis as given in \eqref{eqn:REALSPIN} we can rewrite $\Lambda$ as its singular value decomposition $\Lambda=VDV^{\da}$, where $V$ is a unitary matrix who's columns are the eigenvectors of $\Lambda$ and $D$ is a diagonal matrix containing the eigenvalues of $\Lambda$
\begin{align}
    V &= \bpm \ket{\kappa_{n}},\ket{\kappa_{-n}}, & \dots, &\ket{\kappa_{1}}, & \ket{\kappa_{-1}}\epm,\nn\\
    \text{diag}(D) &= \bpm E_{n}, E_{-n} &\dots,& E_{1},& E_{-1}\epm.
\end{align}
By contracting $V$ with the spinor $\Psi$ we put the Hamiltonian in the form $H=\tilde{\Psi}^{\da}D\tilde{\Psi}$ where $\tilde{\Psi}=V^{\da}\Psi$. The elements of $\tilde{\Psi}$ are the spectral Dirac fermionic operators that diagonalise the Hamiltonian, such that $H=\sum_{n}E_{n}\tilde{a}^{\da}_{n}\tilde{a}_{n}$, and each spectral Dirac fermion operator can be written as a sum of the constituent Dirac fermion operators 
\begin{align}
    \tilde{a}_{n}&= \bra{\kappa_{n}}\Psi = \sum_{\bj k} \big(\alpha_{\bj k \parti}^{\kappa_{i}}\big)^{*} a_{k,\bj}+\big(\alpha_{\bj k \hole}^{\kappa_{i}}\big)^{*} a_{k,\bj}^{\da},\nn\\
    \tilde{a}_{-n}&= \bra{\kappa_{-n}}\Psi = \sum_{\bj k} \alpha_{\bj k \hole}^{\kappa_{i}} a_{k,\bj}+\alpha_{\bj k \parti}^{\kappa_{i}} a_{k,\bj}^{\da}.
\end{align}
where $\alpha_{\bj k \rho}^{\kappa_{i}}$ come from the relevant eigenstates $\ket{\kappa_{i}}$ and we note that $\tilde{a}_{n}^{\da}=\tilde{a}_{-n}$. These spectral Dirac fermion operators can act on a fermionic Fock vacuum $\ket{\varnothing}=\bigotimes_{\bj}\ket{0_{a_{1}}0_{a_{2}}}$, such that
\begin{equation}
    \ket{\psi_{i}} = 2\tilde{a}_{i}^{\da}\ket{\varnothing}= 2\sum_{\bj,k} \alpha_{\bj k \parti}^{\kappa_{i}} a_{k,\bj}^{\da}\ket{\varnothing},
\end{equation}
where the factor of 2 renormalises the state. $\ket{\psi_{i}}$ are the eigenstates of $H$ in the Fock representation.

Another useful representation is the Majorana fermion basis. Via the unitary matrix
\begin{equation}
    U_{\gamma} =\bigoplus_{\bj}\frac{1}{\sqrt{2}}\bpm 1 & 1 & 0 & 0 \\ -i & i & 0 & 0 \\ 0 & 0 & 1 & 1 \\ 0 & 0 & -i & i \epm,
\end{equation}
we can rotate the Hamiltonian into a new basis $H=\Psi^{\da}U_{\gamma}^{\da}U_{\gamma}\Lambda U_{\gamma}^{\da}U_{\gamma}\Psi=\Psi_{\gamma}^{\da}\Lambda_{\gamma}\Psi_{\gamma}$. The elements of $\Psi_{\gamma}$ are the constituent Majorana fermion operators $\gamma_{g,\bj}$, given by
\begin{align}
    \gamma_{1,\bj} &= \frac{a_{1,\bj} + a_{1,\bj}^{\da}}{2}, \quad &\gamma_{2,\bj} = \frac{a_{1,\bj} - a_{1,\bj}^{\da}}{2i},\nn\\
    \gamma_{3,\bj} &= \frac{a_{2,\bj} + a_{2,\bj}^{\da}}{2}, \quad &\gamma_{4,\bj} = \frac{a_{2,\bj} - a_{2,\bj}^{\da}}{2i}.
\end{align}
The eigenstates of $\Lambda_{\gamma}$ are
\begin{equation}
    \ket{\kappa_{n}}_{\gamma}=\sum_{\bj}\sum_{g=1}^{4}\beta_{\bj g}\ket{\bj}\otimes\ket{g}.
\end{equation}

The final corner of the diagram in Fig.~\ref{fig:TRANS} corresponds to the spectral Majorana fermion basis. It is related to the spectral Dirac fermion basis via the unitary rotation $\tilde{U}_{\gamma}$, given by 
\begin{equation}
    \tilde{U}_{\gamma}= \mathbb{I}_{\text{dim}(\Lambda)/2}\otimes\bpm 1 & 1 \\ -i & i \epm,
\end{equation}
where $\mathbb{I}_{\text{dim}(\Lambda)/2}$ is the $\text{dim}(\Lambda)/2$ dimensional identity matrix. The elements of the spinor $\tilde{\Psi}_{\gamma}=\tilde{U}_{\gamma}\tilde{\Psi}$ are the spectral Majorana fermion operators $\tilde{\gamma}_{\tilde{g},n}$, where $\tilde{g}=1,2$. such that
\begin{equation}
    \tilde{\gamma}_{1,n}=\frac{\tilde{a}_{n}+\tilde{a}_{n}^{\da}}{2},\quad\tilde{\gamma}_{2,n}=\frac{\tilde{a}_{n}-\tilde{a}_{n}^{\da}}{2i},
\end{equation}
where each gapless Majorana can be expressed as a linear sum of the $\gamma_{i,\bj}$ operators
\begin{align}\label{eqn:ZEROMAJ}
    \tilde{\gamma}_{1,n}=\frac{1}{\sqrt{2}}\sum_{\bj}\Big[&\Re\big\{\beta_{\bj 1}^{\kappa_{i}}\big\}\gamma_{1,\bj}-\Re\big\{\beta_{\bj 2}^{\kappa_{i}}\big\}\gamma_{2,\bj}\nn\\
                                                          &+\Re\big\{\beta_{\bj 3}^{\kappa_{i}}\big\}\gamma_{3,\bj}-\Re\big\{\beta_{\bj 4}^{\kappa_{i}}\big\}\gamma_{4,\bj}\Big],\nn\\ 
    \tilde{\gamma}_{2,n}=\frac{1}{\sqrt{2}}\sum_{\bj}\Big[&\Im\big\{\beta_{\bj 1}^{\kappa_{i}}\big\}\gamma_{1,\bj}-\Im\big\{\beta_{\bj 2}^{\kappa_{i}}\big\}\gamma_{2,\bj}\nn\\
                                                          &+\Im\big\{\beta_{\bj 3}^{\kappa_{i}}\big\}\gamma_{3,\bj}-\Im\big\{\beta_{\bj 4}^{\kappa_{i}}\big\}\gamma_{4,\bj}\Big].
\end{align}
where $\beta_{\bj g}^{\kappa_{i}}$ come from the relevant eigenstates $\ket{\kappa_{i}}_{\gamma}$.

\end{document}